\documentstyle[epsfig, sprocl]{article}

\def\Journal#1#2#3#4{{#1} {\bf #2}, #3 (#4)}

\def\be{\begin{equation}}
\def\ee{\end{equation}}
\def\bea{\begin{eqnarray}}
\def\eea{\end{eqnarray}}

\begin{document}

\title{On the ``initial'' Angular Momentum of Galaxies}

\author{Tom Abel, Rupert C. Croft, Lars Hernquist}

\address{Harvard Smithsonian Center for Astrophysics, \\
    MA, US--02138 Cambridge\\E-mail: \tt{Hi@TomAbel.com} }


\maketitle\abstracts{Spherical density profiles and specific angular
momentum profiles of Dark Matter halos found in cosmological N--body
simulations have been measured extensively. The distribution of the
total angular momentum of dark matter halos is also used routinely in
semi--analytic modeling of the formation of disk galaxies. However, it
is unclear whether the initial (i.e. at the time the halo is
assembled) angular momentum distributions of baryons is related to
the dark matter at all. Theoretical models for ellipticities in weak
lensing studies often rely on an assumed correlation of the angular
momentum vectors of dark matter and gas in galaxies.  Both of these
assumptions are shown to be in reasonable agreement with high
resolution cosmological smoothed particle hydrodynamical simulations
that follow the dark matter as long as only adiabatic gas physics are
included. However, we argue that in more realistic models of galaxy
formation one expects pressure forces to play a significant role at
``turn--around''. Consequently the torquing force on DM and baryons
will be uncorrelated and their respective angular momenta are not
expected to align. An SPH simulation with ad-hoc feedback is presented
that illustrates these effects. Massive low redshift elliptical
galaxies may be a notable exception where ``light may trace mass''.}

\section{Introduction}
The high frequency of disk galaxies and their exponential profiles
over many scale radii is perhaps one of the most striking global
features of observed galaxies. Theoretical studies investigating the
origin of this angular momentum in these galaxies have a long history
(e.g. Weizs\"acker 1951). In most modern investigations the {\sl
Ansatz} of Mestel (1963) is used where one assumes that any parcel of
fluid to retains its initial specific angular momentum. In
hierarchical structure formation scenarios the luminous parts of
galaxies form from gas that is cooling within a dark matter (DM) halos
that continuously are merging to build larger and larger
objects. These assumption are made also in most recent models of disk
galaxies (e.g. Mo, Mao and White 1998). The typical initial angular
momentum of halos may be estimated from linear theory (Peebles
1969). However, its distribution is typically measured from N--body
simulations that follow the formation and evolution of structure in
the dark matter only.  In the following we derive a rough estimate of
the the magnitude of pressure forces that contribute significantly to
the torquing forces.

\section{Forces and Torques}

The critical density and the total mass, $M$, in a spherical volume
of comoving radius $R$ with an over--density $\delta=\rho/\bar{\rho}_m$ of
the mean matter density ($\bar{\rho}_m=\Omega_m \rho_c$)
are given by,
\begin{eqnarray}
\rho_c &=& \frac{3 H_0^2}{8 \pi G}, \ \ \ 
M =  \delta \frac{4 \pi}{3} R^3 \, \, \Omega_m \rho_c,
\end{eqnarray}
where the symbols have their usual meaning (see e.g. Peebles 1993).
Let $f$ be the fraction of the torquing force from surrounding
structures on this spherical over-density to the radial force of the
sphere. I.e. the torquing force is $F_t = f\, G\, M \rho/r^2$ where
$r$ is a proper radius of the sphere. Here $f$ is necessarily somewhat
smaller than the radial force of its own gravity\footnote{Otherwise
the fluid parcel would not collapse with the sphere.}. To get a rough
idea of what pressure gradient may change the torquing we simply
compare
\begin{eqnarray}
\frac{1}{\rho_b} \frac{\partial P}{\partial r} &=& f\,  \frac{G \,
M}{r^2} 
\end{eqnarray}
Using difference instead of differential, substituting $r = R/(1+z)$
and using $\triangle R$\footnote{This $\delta R$ is taken smaller than
the total radius since for a realistic density distribution most of
the mass is at large radii.}to denote the comoving length scale over which
the pressure gradient is taken we find,
\begin{eqnarray}
 \triangle P &\sim& f\, \frac{3 H_0^4}{16\, G\, \pi}\,
\, \delta^2 \Omega_b\, \Omega_m\, (1+z)^4\, R\, \triangle R.
\end{eqnarray}
Assuming constant density this translates to the very small
temperature gradient of
\begin{eqnarray}
 \triangle T &\sim& f\, \frac{\triangle P\, \mu}{k_B \, \rho_b} = \,
\delta \frac{H_0^2
m_P}{2 k_B} \Omega_m (1+z) R \triangle R, \\
 \triangle T &\approx& 2.6\times 10^3 \, \mbox{K}\, \delta \, \frac{f}{0.1}\, (1+z) R_{Mpc}\, 
\frac{\triangle R_{Mpc}}{0.3}.
\end{eqnarray}

Such small pressure gradients are expected to originate from well
known processes such as ionization fronts from reionization, galactic
outflows driven by black holes, and galactic feedback as required in
semi--analytical models of galaxy formation (e.g. Kauffmann et
al. 2000). The larger the mass of the object the larger the pressure
gradient has to be to alter the angular momentum acquisition of the
gas as compared to the dark matter.

\section{Simulations}

To investigate the relation between angular momentum of the baryons
and the DM we have performed two different cosmological smoothed
particle hydrodynamics simulations using GADGET (Springel, Yoshida,
and White 2000). Both simulations use 128$^3$ for each the DM and the
gas in a periodic volume of 10 comoving Mpc side-length. Both are
evolved from redshift 60 to $z=3$ with initial conditions drawn from a
realization appropriate for a spatially flat universe with
$\Omega_m=0.3$, $\Omega_b h^2 = 0.02$ and $h=0.7$. Both simulations
follow only gravity and adiabatic gas physics. In one run we have
artificially increased abruptly the temperature of the gas to $4\times
10^6$ Kelvin at redshift 7 to mimic a strong feedback case. 
\begin{figure}[th]
\vspace{-.1cm}
\centerline{\psfig{file=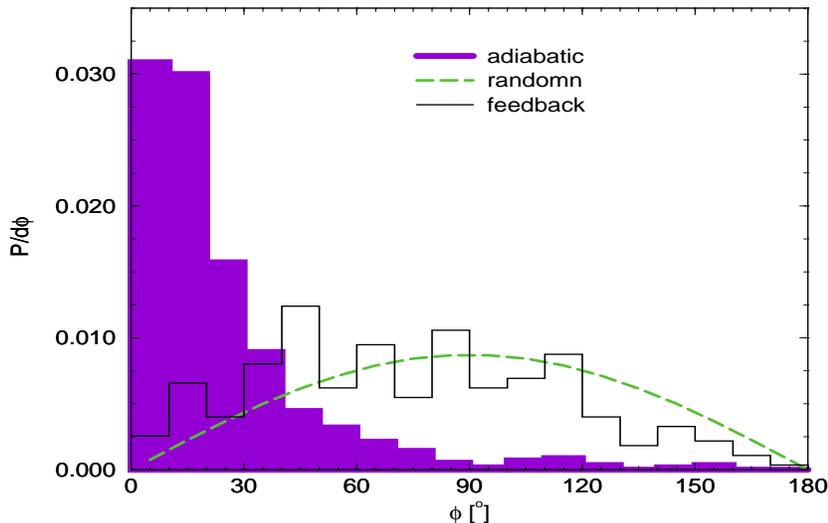, height=7cm,width=11cm}}
\vspace{-.6cm}
\caption{Alignment angle between the total specific angular momentum
vectors of dark matter and gas for purely adiabatic gas physics
(filled histogram) and an ad--hoc feedback model (empty
histogram). The smooth dashed line indicates the expectation for a
purely random distribution.}\label{angle}
\vspace{-.2cm}
\end{figure}
We use HOP (Eisenstein and Hut 1998) to identify centers of halos. The
we go out radially and define a halo as all the mass within the radius
where the over--density in the dark matter corresponds to 200 times
the mean density. Within these halos we then measure the total and
specific angular momentum vectors. In Figure~\ref{angle} we show the
distributions of the alignment angles between the dark matter and the
baryonic component for all halos that have no other overlapping
halo. Clearly for the purely adiabatic case their is remarkable
agreement between the directions of the angular momenta with a median
alignment angle of 20 degrees. However, as soon as pressure forces are
introduces as in the ad--hoc feedback the respective angles become
uncorrelated. The imperfect alignment between the DM and the gas in
the purely adiabatic run comes most likely from non--linear angular
momentum exchange at virialization.  Such non-linear effects may well
be aggravated in simulations which also include the cooling of the
gas. 

\section{Discussion and Conclusions}
In this short contribution we have pointed out that large scale
pressure forces have significant influence on the formation of
galaxies. Until the physical processes acting on the gas in the
intergalactic medium are fully understood it is difficult to asses up
to which mass scale such large scale non-gravitational forces may play
a role. Because of space constraints we could only remark here that
the collisionless and collisional components will exchange angular
momentum during merging events. This will further diminish and
possible correlation of the spins of DM and gas.  Our results have
important implications for models which associate the dark matter
angular momentum vector and their correlations with observed
ellipticities of galaxies (see contributions by Natarayan, Crittenden
and Heavens in this volume).  Since the directions of the angular
momentum of radial shells within halos varies as function of radius
(Moore this volume) and the luminous part of galaxies may be
re--oriented from the torquing their non-spherical parent halos (Moore
and Sellwood, this volume) we are led to conclude that current models
of the formation and evolution of disk galaxies are oversimplified.
However, one interesting exception may be low redshift massive
elliptical galaxies. If most of their stars have formed at early times
in smaller higher redshift objects the shapes and angular momentum
distributions of gas and DM may be strongly correlated. This would be
a natural explanation of the strong lensing observations of Kochaneck
(this volume) which indicate that ``light traces mass'' for these
galaxies.

\end{document}